\documentclass{emulateapj}
\usepackage{threeparttable}
\usepackage{bm}
\usepackage{chngpage}
\usepackage{graphicx}
\usepackage{mathrsfs}

\def\apjl{ApJL}

\shorttitle{Brightening of SGRB 170817A}
\shortauthors{Geng et al.}

\begin{document}

\title{Brightening X-ray/Optical/Radio Emission of GW170817/SGRB 170817A: Results of an Electron-Positron Wind
from the Central Engine?}

\author{Jin-Jun Geng\altaffilmark{1,2,3}, Zi-Gao Dai\altaffilmark{1,2}, Yong-Feng Huang\altaffilmark{1,2},
Xue-Feng Wu\altaffilmark{4}, Long-Biao Li\altaffilmark{1,2}, Bing Li\altaffilmark{1,2}, Yan-Zhi Meng\altaffilmark{4}}

\altaffiltext{1}{School of Astronomy and Space Science, Nanjing University, Nanjing 210023, China; gengjinjun@nju.edu.cn, dzg@nju.edu.cn, hyf@nju.edu.cn}
\altaffiltext{2}{Key Laboratory of Modern Astronomy and Astrophysics (Nanjing University), Ministry of Education, Nanjing 210023, China}
\altaffiltext{3}{Department of Physics, Nanjing University, Nanjing 210093, China}
\altaffiltext{4}{Purple Mountain Observatory, Chinese Academy of Sciences, Nanjing 210008, China; xfwu@pmo.ac.cn}

\begin{abstract}
Recent follow-up observations of the binary neutron star (NS) merging event GW170817/SGRB 170817A reveal that
its X-ray/optical/radio emissions are brightening continuously up to $\sim 100$ days post-merger.
This late-time brightening is unexpected from the kilonova model or the off-axis top-hat jet model for gamma-ray burst afterglows.
In this paper, by assuming that the merger remnant is a long-lived NS, we propose that the interaction between an
electron-positron-pair ($e^+e^-$) wind from the central NS and the jet could produce a
long-lived reverse shock, from which a new emission component would rise and can interpret current observations well.
The magnetic-field-induced ellipticity of the NS is taken to be $4 \times 10^{-5}$ in our modeling, so that the braking of
the NS is mainly through the gravitational wave (GW) radiation rather than the magnetic dipole radiation, and the emission luminosity
at early times would not exceed the observational limits.
In our scenario, since the peak time of the brightening is roughly equal to the spin-down time
scale of the NS, the accurate peak time may help constrain the ellipticity of the remnant NS.
We suggest that radio polarization observations of the brightening would help to distinguish
our scenario from other scenarios. Future observations on a large sample of short gamma-ray burst afterglows or detections of GW signals
from merger remnants would test our scenario.
\end{abstract}


\section{INTRODUCTION}

Recently, the coincident detection of a gravitational wave (GW) event GW170817~\citep{Abbott17a} and its electromagnetic counterparts
(i.e., a short gamma-ray burst SGRB 170817A and a kilonova,~\citealt{Goldstein17,Arcavi17,Pian17,Smartt17,Kasen17,Kasliwal17,Drout17})
confirmed the hypothesis that
binary neutron star (NS) mergers are at least the progenitors of some SGRBs.
Early temporal and spectral observations in optical bands (within $\sim 15$ days) can be interpreted as
the quasi-thermal radiation from a kilonova~\citep{Li98,Rosswog05,Metzger10,Metzger17}.
{\it Chandra} observations showed an X-ray source at location coincident with the kilonova transient
at $\sim 9$ days after the burst~\citep{Troja17,Haggard17}.
Continued monitoring revealed that the X-ray source brightened with time till $15.1$ days post-merger \citep{Margutti17}.
Most recently, this X-ray emission is found to be brightening according to the deep {\it Chandra} observations
at 109.2 days post-burst~\citep{Ruan18}.
In the radio band, it is showed that the radio emission is also brightening slowly up to 93 days post-burst~\citep{Hallinan17,Mooley17}.
Furthermore, it is found that the X-ray emission brightens at a similar rate as the radio emission,
indicating that they share a common origin~\citep{Ruan18}.
Late-time optical emission detected by {\it HST}~\citep{Lyman18} at $\sim 100$ days also supports that the optical brightening is from the same origin.

The physical origin of the prompt emission and afterglow of SGRB 170817A has not been revealed yet.
A uniform jet with a sharp edge (also called a top-hat jet) is usually considered to be responsible
for the GRB prompt emission previously \citep[e.g.,][]{Panaitescu99,Meszaros99}.
The low isotropic luminosity ($\sim $ $10^{47}$ erg s$^{-1}$) of the prompt emission of SGRB 170817A
suggests that the GRB jet should be seen by an off-axis observer in this case~\citep{Ioka17,Granot17a}.
However, the contradiction of the ratios of typical off-axis to on-axis photon energy and isotropic-equivalent energy
does not support that the prompt emission originates from such a top-hat jet~\citep{Kasliwal17,Granot17b}.
Moreover, the recently-observed brightening X-ray/optical/radio afterglow also rules out
the simple off-axis top-hat jet origin since it is not in agreement with the data.
Alternative explanations, such as a structured off-axis jet~\citep{Kathirgamaraju18,Xiao17,Meng18},
or the Thomson-scattered emission with a typical SGRB jet~\citep{Kisaka17},
or the breakout of a mildly relativistic wide-angle cocoon~\citep{Nakar10,Gottlieb17}, 
were further suggested to be responsible for the prompt emission.
The brightening afterglow is shown to be well interpreted by
a structured outflow with a highly relativistic core~\citep{Lazzati17c,Lyman18,Margutti18},
which supports a jet-cocoon system produced in the NS merger
~\citep{Nagakura14,Murguia14,Nakar17,Lazzati17a,Lazzati17b,Wang17}.

Except for the scenarios mentioned above, it is still possible that the brightening is caused by a continued injection of energy from the central engine
into the external jet~\citep{Pooley17,Li18}. The energy injection has been frequently invoked in interpreting many nonstandard
afterglow behaviors, such as the X-ray plateau~\citep[e.g.,][]{Dai98a,Dai04,YuDai07,Yu07,Rowlinson10,Bucciantini12,Gompertz13,
Rowlinson13,Zhang13,Fan13,Lv15} and the optical
rebrightening~\citep[e.g.,][]{Dai98b,Liu10,Xu10,Geng13,Laskar15,Yu15,Geng16,GengH16,ZhangQ16}.
Moreover, the energy injection is a natural result of late activities of the central engine.
If the remnant of the GRB progenitor is an NS, inspired by observations/theories of pulsar wind nebulae~\citep{Kargaltsev08,Porth17,Reynolds17},
it is suggested that continuous ultrarelativistic electron-positron-pair ($e^+e^-$) wind would flow into the prior external shock~\citep{Rees74,Lyubarsky01,Dai04,Porth13,Metzger14}.
Otherwise, if the remnant is a black hole, the injected flow may be Poynting-flux dominated~\citep{Blandford77}.

After the binary NS merger, a supramassive or hypermassive remnant NS may survive from the
merger for a sufficiently long time~\citep{Faber12,Piro17}.
For a stiff equation of state of supranuclear matter, the merger remnant may be a long-lived
millisecond pulsar (e.g., an NS or even a quark star) \citep{Dai98b,Dai06}.
On the other hand, upper limits placed on the strength of GW emission by LIGO
cannot definitively rule out the existence of a long-lived postmerger NS~\citep{Abbott17a}.
Therefore, in the case of SGRB 170817A, it is still reasonable to consider that the remnant may be a long-lived NS.
Following the model of an interaction between the $e^+e^-$ wind and the relativistic external shock \citep{Dai04}
and motivated by the success of this model
in explaining the optical rebrightening in some GRB afterglows~\citep{Geng16}, we here propose that the brightening
of X-ray/optical/radio afterglow in SGRB 170817A could also be the result of $e^+e^-$ wind injected into the external shock.
The reverse shock (RS) produced during the interaction between the $e^+e^-$ wind and the external shock
would heat the cold ultrarelativistic $e^+e^-$ plasma efficiently, which would lead to the late-time brightening.

This paper is organized as follows. We briefly present the dynamic method used in our work and the formulae for
calculating the radiation in Section 2. In Section 3, we show that our
scenario works well to explain the afterglow brightening of SGRB 170817A.
The rationality and relevant applications of our scenario are further discussed in Section 4.
Finally, in Section 5, we summarize our results.

\section{Hydrodynamics and Radiation}

When a relativistic jet propagates into the circum-burst
medium, a forward shock (FS) will develop. Assuming the central remnant is an NS and a
continuous wind is ejected into the FS, then a reverse shock (RS) would form.
Two shocks separate the system into four regions \citep{Dai04}: (1) the unshocked ISM, (2)
the shocked ISM, (3) the shocked wind, and (4) the unshocked wind.
The dynamics of such an FS-RS system can be analytically solved under some assumptions \citep{Dai04} and numerically solved by the mechanical
method~\citep{Beloborodov06,Uhm11,Uhm12} or the energy conservation method~\citep{Huang99,Huang00,Geng14,Geng16}.

Assuming that the bulk Lorentz factor of the unshocked particle wind is $\Gamma_4$,
the particle density in the comoving frame of Region 4 at radius $r$ is then
\begin{equation}
n_4^{\prime} = \frac{L_{\rm w}}{4 \pi r^2 \Gamma_4^2 m_{\rm e} c^3 (1 + \sigma)},
\end{equation}
where $m_{\rm e}$ is the mass of an electron, $c$ is the speed of light and $\sigma$ is the
magnetization parameter of the wind. The comoving magnetic field of the wind is
$B_4^{\prime} = (4 \pi n_4^{\prime} m_{\rm e} c^2 \sigma)^{1/2}$ accordingly.
Hereafter, quantities of Region ``$i$'' are denoted by subscripts ``$i$'' respectively.
The number density and the internal energy density of Region 3, $n_3^{\prime}$ and $e_3^{\prime}$,
can be obtained by the shock jump conditions (for the situation of $\sigma \ne 0$, \citealt{Coroniti90,Zhang05,Mao10,Liu16}) as
$e_3^{\prime} = (\Gamma_{34}-1) f_a n_3^{\prime} m_{\rm e} c^2$, $n_3^{\prime} = (4 \Gamma_{34} + 3) f_b n_4^{\prime}$,
where $\Gamma_{34}$ is the relative Lorentz factor of Region 3 measured in Region 4,
the detailed expressions of $f_a$ and $f_b$ could be found in \cite{Zhang05}.
For the FS, the shock jump conditions (for $\sigma = 0$) give the number density and the internal energy density of Region 2
as $e_2^{\prime} = (\Gamma_2-1) n_2^{\prime} m_{\rm p} c^2$, $n_2^{\prime} = (4 \Gamma_2 + 3) n_1$,
where $m_{\rm p}$ is the mass of proton and $\Gamma_2$ is the bulk Lorentz factor of the FS.

As usual, we consider synchrotron radiation from electrons shocked by the FS and the RS.
The energy distribution of the shocked electrons is assumed to be a power-law as
$dN_{{\rm e},i}^{\prime}/d\gamma_{{\rm e},i}^{\prime} \propto \gamma_{{\rm e},i}^{\prime -p_i}
(\gamma_{{\rm m},i}^{\prime} \leq \gamma_{{\rm e},i}^{\prime} \leq \gamma_{{\rm M},i}^{\prime})$,
where $\gamma_{e,i}^{\prime}$ is the Lorentz factor of electrons in Region $i$,
$\gamma_{{\rm m},i}^{\prime}$ is the minimum Lorentz factor, $\gamma_{{\rm M},i}^{\prime}$ is the maximum Lorentz factor,
and $p_i$ is the spectral index.
The minimum Lorentz factor is calculated as
\begin{equation}
\gamma_{{\rm m},i}^{\prime} = \zeta_i \xi_{{\rm e},i} \frac{p_i-2}{p_i-1} (\hat{\Gamma}_i-1) + 1,
\end{equation}
where $\zeta_2 = m_{\rm p}/m_{\rm e}$, $\zeta_3 = 1$, $\hat{\Gamma}_2 = \Gamma_2$, and $\hat{\Gamma}_3 = \Gamma_{34}$,
$\xi_{{\rm e},i}$ is the energy equipartition parameter for shocked electrons.
The comoving-frame magnetic field of the shocked region is
$B_i^{\prime} = (8 \pi e_i^{\prime} \xi_{B,i})^{1/2}$, where $\xi_{B,i}$ is the energy equipartition parameters for
magnetic fields. For $i = 3$, one should note the underlying natural condition of $\xi_{{\rm e},3} + \xi_{B,3} = 1$
and an additional magnetic field component from the upstream (i.e.,
$(4 \Gamma_{34} + 3) f_b B_4^{\prime}$ for a perpendicular shock) should be included in $B_3^{\prime}$
\footnote{For $\sigma \le 1$, $\sigma$ are equivalent to $\xi_{B,3}$ in view of
the ratio of $B_3^{\prime}$ to $e_3^{\prime}$. We simply take $\sigma = \xi_{B,3}$ in our calculations
since they are degenerate in the resulted flux density.}.
Due to the radiation loss, the actual electron distribution would also be
characterized by the cooling Lorentz factor $\gamma_{{\rm c},i}^{\prime}$, which is given by
\begin{equation}
\gamma_{{\rm c},i}^{\prime} = \frac{6 \pi m_{\rm e} c (1+z)}{\sigma_{T} B_i^{\prime 2} (\Gamma_i + \sqrt{\Gamma_i^2-1})
t_{\rm obs}},
\end{equation}
where $\sigma_T$ is the Thomson cross section, $t_{\rm obs}$ is the observed time
and $z$ is the redshift of the source.

For an off-axis observer, the observed flux density can be calculated by
integrating the emission from a series of rings centering at the line-of-sight (LOS), i.e.,
\begin{equation}
F_{\nu,i} = \frac{1+z}{4 \pi D_L^2} \int_{\theta_{-}}^{\theta_{+}} \mathcal{D}^3 \sin \theta d \theta \int_{-\Delta \phi}^{\Delta \phi}
P_{\nu^{\prime},i}^{\prime} d \phi,
\end{equation}
where $P^{\prime}_{\nu^{\prime},i} = \int \frac{d N_{{\rm e},i}^{\prime}}{d \gamma_{{\rm e},i}^{\prime}} p^{\prime}_{\nu^{\prime}}
d \gamma_{{\rm e},i}^{\prime}$,
$\frac{d N_{{\rm e},i}^{\prime}}{d \gamma_{{\rm e},i}^{\prime}}$ is the equivalent isotropic number distribution of electrons in the emitting shell,
$p^{\prime}_{\nu^{\prime}}$ is the synchrotron emission power at frequency $\nu^{\prime}$
for an electron of Lorentz factor $\gamma_{{\rm e},i}^{\prime}$,
$\mathcal{D} = [\Gamma_2 (1-\beta_2 \cos \theta)]^{-1}$ is the Doppler factor,
and $D_L$ is the luminosity distance of the burst.
The corresponding integral limits are obtained according to virtue of spherical geometry~\citep{Wu05},
$\Delta \phi = \arccos \left( \frac{\cos \theta_j - \cos \theta_V \cos \theta}{\sin \theta_V \sin \theta} \right)$,
for case of $\left| \theta_j - \theta_V \right| < \theta <  \theta_j + \theta_V$ here.
In our calculations, for the flux at $t_{\rm obs}$, the integration is performed over the equal arrival time surface
(EATS, \citealt{Waxman97,Granot99,Huang07,Geng17}), which is determined by
\begin{equation}
t_{\rm obs} = (1+z) \int_0^{R_{\theta}} \frac{1-\beta_i \cos \theta}{\beta_i c} dr \equiv {\rm const},
\end{equation}
from which $R_{\theta}$ (the radius for postion at $\theta$ on the surface) can be derived.

\section{Modeling The Afterglow}

If a wind from the NS is due to magnetic dipole radiation (MDR), its luminosity is \citep{Shapiro83,YuDai17}
\begin{equation}
L_{\rm w} \simeq 9.6 \times 10^{42} B_{\rm NS,12}^2 R_{\rm NS,6}^6 P_{{\rm NS},-3}^{-4} \left(1+\frac{t_{\rm obs}}{T_{\rm sd}}\right)^{-\alpha} \rm{erg}~\rm{s}^{-1},
\end{equation}
where $B_{\rm NS}$, $R_{\rm NS}$, $P_{\rm NS}$, $T_{\rm sd}$ and $\alpha$
are the strength of the polar magnetic field, the radius, the initial spin period,
the spin-down timescale and the decay index of the NS.
$T_{\rm sd}$ and $\alpha$ are further determined by considering
whether the spin-down of the NS is mainly due to the MDR or the GW radiation.
The convention $Q_x = Q/10^x$ in cgs units is adopted hereafter.
Within the scenario of $e^+e^-$ injection, it was revealed that the peak time of the
brightening X-ray afterglow are roughly determined by $T_{\rm sd}$~\citep{Geng16}.
As a consequence, as required by the temporal observational data, we have
\begin{equation}
T_{\rm sd} \ge 110~\mathrm{days} \approx 9.5 \times 10^6~\mathrm{s}.
\label{eq:Tsd}
\end{equation}
If the spin-down of the NS is due to MDR, then
\begin{equation}
T_{\rm sd,MDR} = 2 \times 10^9 I_{45} B_{\rm NS,12}^{-2} R_{\rm NS,6}^{-6} P_{{\rm NS},-3}^2~\mathrm{s},
\label{eq:Tsd_MDR}
\end{equation}
where $I$ is the moment of inertia of the NS.
Combining Equations (\ref{eq:Tsd}) and (\ref{eq:Tsd_MDR}) gives
\begin{equation}
B_{\rm NS} \le 1.5 \times 10^{13} I_{45}^{1/2} R_{\rm NS,6}^{-3} P_{{\rm NS},-3} ~{\rm G},
\end{equation}
where typical values of $I = 10^{45}$~g~cm$^2$, $R_{\rm NS} = 10^6$~cm,
and $P_{\rm NS} = 10^{-3}$~s are taken.

On the other hand, GW radiation may also brake a newly born NS efficiently.
The GW emission mechanism from the remnant NS may be due to
magnetic-field-induced ellipticities \citep{Bonazzola96,Palomba01,Cutler02,Dall09}, unstable bar modes
~\citep{Lai95,Corsi09} or unstable r-modes oscillations \citep{Andersson98,Lindblom98,Dai16}.
Here, we consider the first case to perform some representative estimates.
The spin-down timescale due to an ellipticity of $\epsilon$ can be expressed as \citep{Shapiro83,Usov92,CutlerJones02}
\begin{equation}
T_{\rm sd,GW} = \frac{5 c^5 P_{\rm NS}^4}{2048 \pi^4 G I \epsilon^2} = 9.1 \times 10^7 \epsilon_{-5}^{-2} I_{45}^{-1}
P_{{\rm NS},-3}^4~\mathrm{s},
\label{eq:Tsd_GW}
\end{equation}
where $G$ is the gravitational constant.
Combining Equations (\ref{eq:Tsd}) and (\ref{eq:Tsd_GW}), we derive that
\begin{equation}
\epsilon \le 3.1 \times 10^{-5} I_{45}^{-1/2} P_{{\rm NS},-3}^2.
\end{equation}

How the NS spins down is uncertain, so the
two cases with different braking mechanisms mentioned above are taken into account in our modeling.
Due to the substantial angular momentum of the initial binary,
an initial spin period of $P_{\rm NS} \sim 10^{-3}$~s close to the centrifugal break-up limit is expected
for the natal NS~\citep[e.g.,][]{Rezzolla11,Giacomazzo13}.
We thus fix $P_{\rm NS}$ to be 1~ms for all our modeling.
Model parameters making a good match to the data are given in Table~\ref{Table:para}
and the corresponding light curves are shown in Figure~1.
The values for parameters like $\theta_j$, $\theta_V$, $p_2$, $p_3$ and $n_1$ adopted
are consistent with those in other works~\citep{Lazzati17c,Lyman18,Margutti18}.
From Figure~1, it is seen that the multi-wavelength brightening
is attributed to the emission from Region 3.
The synchrotron frequency of $\gamma_{\mathrm{m},3}$ ($\nu_{\mathrm{m},3}$)
is below the radio band (3 GHz), while the cooling frequency $\nu_{\mathrm{c},3}$
is above the X-ray band (1 keV) in our calculations,
which is consistent with the almost same temporal indices for multi-wavelength brightening.
In our modeling, we have supposed that the peak time (still unknown) of afterglows is roughly around $150$ days.
Since $T_{\rm sd}$ is sensitive to $B_{\rm NS}$ or $\epsilon$ in two scenarios respectively,
a smaller $B_{\rm NS}$ or $\epsilon$ would lead to a later peak time.
The true peak time from future data would help us to determine the realistic values for
$B_{\rm NS}$ and $\epsilon$, together with slight adjustment of other parameters.

\begin{table}
\centering
\caption{Parameters used in the modeling of the afterglow of SGRB 170817A.\label{Table:para}
In our calculations, we have considered two different models, in which the dominant
braking mechanism for the spinning down of the pulsar is MDR or GW radiation, respectively.}
\begin{threeparttable}
\begin{tabular}{lcc}
\toprule Parameters & MDR &  GW radiation \\

$\theta_j$     & $11^{\circ}$  & $10^{\circ}$ \\
$\theta_V$     & $20^{\circ}$  & $20^{\circ}$ \\
$E_{\rm K,iso}$\tnote{a}~($10^{50}$~erg) &  5 &  2 \\
$\Gamma_0$\tnote{b}            &  100    &  100    \\
$p_2$                          &  2.12   &  2.15   \\
$\xi_{{\rm e},2}$       &  0.012  &  0.05   \\
$\xi_{B,2}$             &  0.003  &  0.002  \\
$n_1$~(cm$^{-3}$)       &  $8 \times 10^{-4}$  & $8 \times 10^{-4}$ \\
\hline
$B_{\rm NS}$~(G)        &  $1.3 \times 10^{13}$ &  $10^{12}$ \\
$\Gamma_4$\tnote{c}     &  $10^4$    &  $10^4$    \\
$p_3$                   &  2.17      &  2.17      \\
$\xi_{B,3}$             &  $2 \times 10^{-7}$   &  0.02          \\
$\xi_{{\rm e},3}$       &  $1-\xi_{B,3}$        &  $1-\xi_{B,3}$ \\
$\sigma$                &  $2 \times 10^{-7}$   &  0.02          \\
\hline
$\epsilon$              &     0      & $4.4 \times 10^{-5}$ \\
\hline
\end{tabular}

\begin{tablenotes}
\item[a] $E_{\rm K,iso}$ is the initial isotropic kinetic energy of the GRB jet ejecta.
\item[b] $\Gamma_0$ is the initial Lorentz factor of the GRB jet ejecta.
\item[c] $\Gamma_4$ is typically in range of $10^4-10^7$ according to observations
and theories of pulsar wind nebula~\citep[e.g.,][]{Rees74,Lyubarsky01,Bucciantini11}.
\end{tablenotes}
\end{threeparttable}
\end{table}

\begin{figure}
   \begin{center}
   \includegraphics[scale=0.4]{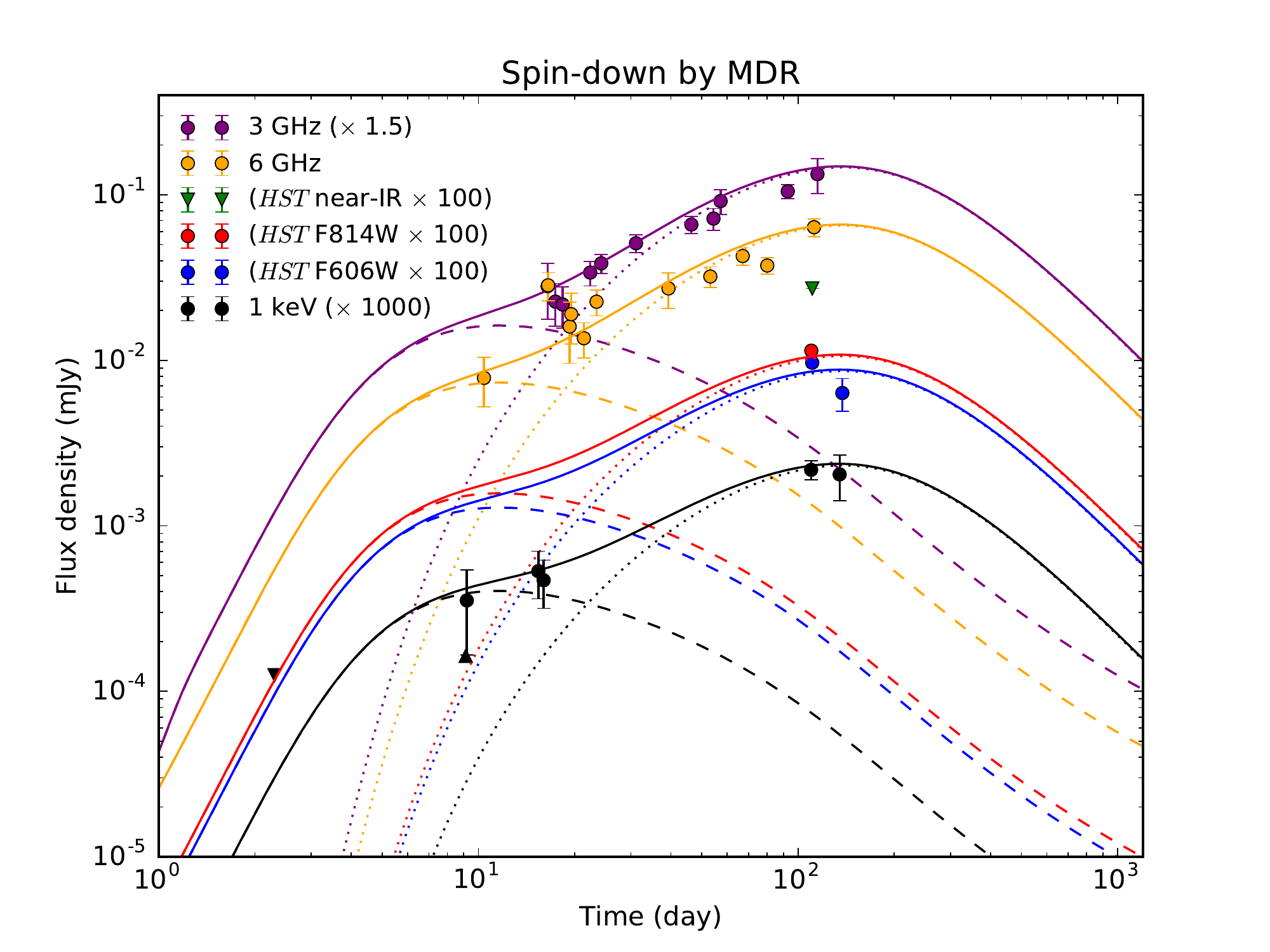} \\
   \includegraphics[scale=0.4]{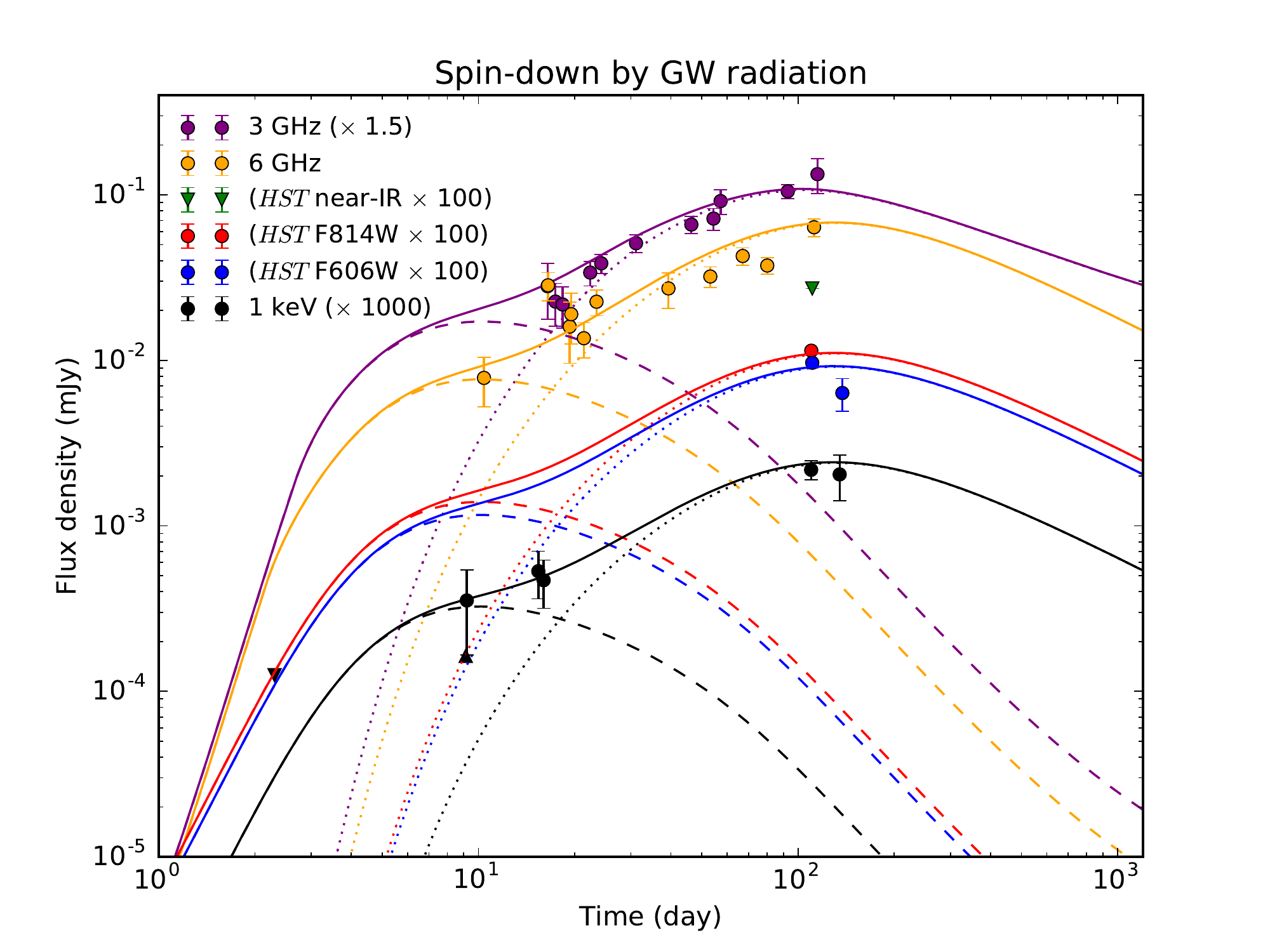}
   \caption{Fitting to the brightening of X-ray/optical/radio afterglow of SGRB 170817A.
   In the upper panel, the NS is braked mainly by the MDR, while the NS is braked
   mainly by the GW radiation in the lower panel.
   In each panel, dashed, dotted and solid lines represent the flux from the FS (Region 2),
   the RS (Region 3) and the total flux respectively.
   Note that the flux densities in bands of 1~keV and 3 GHz have been multiplied
   by different constants for viewing purposes.
   The radio data are compiled from~\cite{Alexander17},~\cite{Hallinan17},~\cite{Kim17},
   \cite{Mooley17}, and \cite{Margutti18},
   the X-ray data are taken from~\cite{Ruan18},~\cite{Lazzati17c}, and~\cite{DAvanzo18},
   and the optical (in {\it HST} optical bands of F814W and F606W) data are taken from
   \cite{Lyman18}, and~\cite{Margutti18}.
   The flux upper limit in {\it HST} near-IR band is shown as a green triangle.}
   \end{center}
   \label{Fig:plot1}
\end{figure}

\section{Discussion}

When the spin-down of the NS is dominated by the MDR, our modeling requires that
$\xi_{B,3}$ and $\sigma$ should be extremely low.
This low radiation efficiency is to overcome the high peak flux density
($F_{\nu,\mathrm{max},3} \propto \xi_{B,3}^{1/2} L_{\rm w}^{3/2} \propto \xi_{B,3}^{1/2} B_{\rm NS}^{3}$)
caused by the relatively large $B_{\rm NS}$.
Some other authors also found that the high spin-down luminosity from a magnetar
remnant is much higher than the bolometric luminosity of the kilonova,
which has been claimed to disfavor a long-lived magnetar remnant.
This magnetar scenario is not preferred in comparison with the GW spin-down scenario too.
In the case that the spin-down of the NS is dominated by
the GW radiation, the parameter $B_{\rm NS}$ is free to a fix $T_{\rm sd}$ so that
$\xi_{B,3}$ and $\sigma$ could be adjusted in a more plausible range under small $B_{\rm NS}$.
But an ellipticity of $10^{-5}$ for the NS is needed.
This ellipticity could be caused by an internal toroidal magnetic field of
$B_{\rm tor} \approx 2 \times 10^{15} \epsilon_{-5}^{1/2}$~G~\citep{Bonazzola96,Cutler02,Dall09,Mastrano11,Gao17}.
The strength of the polar magnetic field used in our fitting
is $B_{\rm NS} = 10^{12}$~G, which is much smaller than $B_{\rm tor}$.
However, the analyses on the stability of rotating NS with different configurations of the poloidal (surface)
and the internal toroidal component by~\cite{Akgun13} and \cite{Dall15} show that
the magnetic configuration of the NS is stable, if the maximum toroidal field strength meets
\begin{equation}
B_{\rm tor,max} \le 3 \times 10^{15} B_{\mathrm{NS},12}^{1/2}~\mathrm{G}.
\end{equation}
The $B_{\rm tor}$ required for the ellipticity in our result could be still below the $B_{\rm tor,max}$.
Moreover, our adopted $\epsilon$ of less than a few times $10^{-5}$ is
consistent with current upper limits on the GW emission~\citep{Abbott17b}.
It is well below the LIGO detectability, as would be any higher, yet plausible value.

It was argued that high X-ray emission from the central NS with $B_{\rm NS} \ge 10^{12}$~G
could be detected in the early afterglow~\citep[e.g.,][]{Zhang13,Sun17}.
This early high X-ray emission may be in contrary with the upper limits of X-ray flux for SGRB 170817A
\citep{Margutti18}.
Indeed, the wind luminosity of $L_{\rm w} = 9.6 \times 10^{42}~\mathrm{erg}~\mathrm{s}^{-1}$ ($B_{\rm NS} = 10^{12}$~G)
is higher than the X-ray upper limit of $7 \times 10^{38}$~erg~s$^{-1}$ at $\sim 2.2$~days \citep{Troja17}.
However, the radiation efficiency of converting wind luminosity to X-ray luminosity ($\eta_X$) needs to be taken into account
when we are using this upper limit to constrain the parameters of the NS.
If the wind from the NS is dissipated via some mechanisms such as gradual magnetic reconnection \citep{Spruit01,Drenkhahn02,Beniamini17},
$\eta_X$ will increase monotonously with $L_{\rm w}$
for a specific $\Gamma_4$ and will be inversely dependent on $\Gamma_4$ (see Figure 4 of \citealt{Xiao17a}).
For the case of $\Gamma_{4} = 10^4$ and $B_{\rm NS} = 10^{12}$~G adopted
in our paper, $\eta_X$ is less than $10^{-4}$ according to \cite{Xiao17a},
which ensures that the X-ray emission (if exists) from the NS wind would not exceed the observational limit.
On the other hand, the energy of the NS wind may be absorbed by the merger ejecta,
so the NS wind is a further energy source for the kilonova in addition to the radioactivity.
In the scenario proposed by~\cite{Metzger14}, a nebula of $e^+e^-$ and non-thermal photons is
produced from the dissipation of the NS wind behind the ejecta.
A fraction of the nebula energy would be absorbed by the ejecta (non-thermal photons get thermalized via
their interaction with the ejecta wall), and a range of $0.01-0.1$ is generally expected for the absorption efficiency~\citep{Metzger14,YuDai17}.
The remaining fraction of the nebula energy will be lost by the $PdV$ work, which affects the evolution of the ejecta radius.
Considering the absorption efficiency of $0.01-0.1$ for the NS wind here, 
the resulted power is lower than the bolometric luminosity of the kilonova ($\sim 5 \times 10^{41}~\mathrm{erg}~\mathrm{s}^{-1}$,~\citealt{Cowperthwaite17}). So the NS wind would not affect the temporal behaviour 
of the kilonova significantly.

In our calculations, we have only considered the $e^+e^-$ wind that is injected into
the GRB jet (along the jet direction).
Observations of more slowly rotating pulsars (e.g., Crab) indicate that the jet-and-torus morphology
is common for the pulsar wind nebula~\citep[e.g.,][]{Kargaltsev08,Reynolds17}.
If the $e^+e^-$ wind is isotropic or toroidally dominated (such as the Crab nebula), the effect of
its injection into the kilonova ejecta may be more significant.
However, simulations show that it is still possible that the $e^+e^-$ wind
would be dominated in the jet direction when the magnetic obliquity angle is small (see Figure~5 of \citealt{Porth17}).
At the same time, we are still lacking of the knowledge on the preferred wind direction
of a newly born NS studied here. So our calculations still make sense as long as there is an $e^+e^-$ wind
injected in the jet direction.

In both scenarios, we can see that the resulted multi-wavelength afterglows are observable
within $\sim 1000$ days, which could help to test our modeling in the future.
However, one should notice that, due to the spin-down, the NS, if supramassive,
may collapse into a black hole at some late time $\ge T_{\rm sd}$.
After the collapse, the $e^+e^-$ wind injection may be terminated and a steeper decay
in the lightcurves after the peak is expected.
We have not included this possibility in our current calculations.
An abrupt decay signature (if exist) in the late observational data would correspond to this possibility.

It may be difficult to distinguish our model from the structured jet/cocoon model merely
based on the lightcurves. Both our model and the structured jet/cocoon model can well interpret the
data in the brightening phase till now. For the upcoming decay phase, the decay index
of the optical lightcurves ranges from $-0.5$ to $-1.1$ in the structured jet model~\footnote{
However, it should be noticed that the decay index is $\sim -1.2$ for the cocoon model
and $\sim -2.5$ for the jet model as claimed in~\cite{Troja18}.},
depending on different $n_1$ used~\citep{Margutti18}.
The decay index in our model is $\sim -0.7$, which does not deviate much from the former.
Nevertheless, observations of the polarization during the brightening phase may help to judge which model is preferred.
According to~\cite{Lan16}, the magnetic field frozen in the relativistic wind
may be large-scale ordered and synchrotron radiation from the shocked wind region
should be highly polarized. While the magnetic field of the emission region in the structured jet/cocoon
model may be random, a relatively low polarization degree would be expected.
Thus we suggest radio polarimeter facilities should be used to detect the polarization
evolution of the radio afterglow of SGRB 170817A.

\section{Conclusions}

As shown by deep {\it Chandra}, {\it HST}, and VLA observations of GW170817,
the multi-wavelength brightening has ruled out the model of a simple
off-axis top-hat jet. Here we have proposed that the $e^+e^-$ wind injection from a central NS
into a top-hat jet could yet provide an explanation for current observations.
The brightening is attributed to the synchrotron emission from electrons accelerated by the RS
in our scenario. The peak time observed in the future would mark the spin-down time of the central NS.

In our modeling, we invoked the GW radiation as the main braking mechanism of the NS.
An ellipticity of $4.4 \times 10^{-5}$ and a polar magnetic field of $10^{12}$~G
are required to match current data. Other relevant parameters are generally consistent
with those adopted in other works. Current observations cannot distinguish our scenario
from other models such as the cocoon model and the structured jet model~\citep{Mooley17,Gottlieb17,Lazzati17c,Lyman18,Margutti17}.
Further detailed observations of lightcurves and polarization evolution may help to discriminate these models.

The observations on afterglows of long GRBs have shown some features
that are beyond the standard model, such as the X-ray plateau and optical rebrightening.
Lessons from modeling these (on-axis) special afterglows tell us that it is still uncertain
whether the optical rebrightening is caused by the late activities of a central engine
or detailed characteristics of the jet (like the two-component jet model, a kind of simple structured jet).
This situation may also exist for SGRB afterglows, as in the case of SGRB 170817A here.
No such deep multi-wavelength follow-ups of SGRBs has been carried out
before SGRB 170817A, hence the sample of SGRBs with high quality data for late-time afterglows
is still rare. In the future, there are two potential ways to distinguish our $e^+e^-$ scenario from
the structured jet scenario. The first is an indirect way.
For late-time afterglows of an on-axis jet from a double NS merger, the lightcurves
are thought to gradually decline in the scenarios of the structured jet or the jet-cocoon system.
This gradual decline is a natural result of the smooth distribution of $\Gamma$ and $E_{\rm iso}$
from simulations when the jet is viewed on axis.
However, for the $e^+e^-$ wind injection scenario, the flux from the RS should be able
to exceed that from the FS. Therefore, significant optical rebrightening at late times may occur
in the sample of on-axis afterglows from double NS merger.
In fact, the clear optical rebrightening has been observed in SGRB 090426~\citep{Nicuesa11,Geng16},
although some authors argued that GRB 090426 may be associated with a collapsar event, rather than the
merger of double NSs~\citep{Antonelli09,Levesque10,Xin11}.
Secondly, the post-merger remnant could be confirmed by searching the GW signals from
the remnant. The identification of a central long-lived NS or others would support or rule out our scenario.
It could be done when the advanced GW detectors reach their designed sensitivity or with the next-generation detectors~\citep{Abbott17b}.

\acknowledgments
We thank the anonymous referee for valuable suggestions.
We also thank Xiang-Yu Wang for helpful discussions.
This work is supported by the National Basic Research Program of China (``973'' Program, Grant
No. 2014CB845800), the National Key Research and Development
Program of China (grant no. 2017YFA0402600), the National Postdoctoral Program for Innovative Talents
(Grant No. BX201700115), the China Postdoctoral Science Foundation funded project
(Grant No. 2017M620199), the National Natural Science Foundation of China (Grants No.
11473012, 11433009, 11673068, 11573014 and 11725314), and by the Strategic Priority Research Program of the Chinese Academy of Sciences
``Multi-waveband Gravitational Wave Universe''  (Grant No. XDB23040000).

\end{document}